\documentclass[9pt]{IEEEtran}

\usepackage{lscape}
\usepackage{graphicx,caption,subcaption}
\usepackage{amsmath}
\usepackage{mathtools}
\usepackage{amssymb}
\usepackage{enumitem}
\usepackage{epstopdf} 
\usepackage{tikz}
\usepackage{setspace} 
\usepackage{tabularx}
\usepackage{booktabs}
\usetikzlibrary{trees}
\usetikzlibrary{fit}
\usepackage{amsthm} 
\newtheorem{definition}{Definition}[section]
\usepackage{hyperref}
\hypersetup{
    colorlinks,
    citecolor=black,
    filecolor=black,
    linkcolor=black,
    urlcolor=black
}
\usepackage[mathscr]{euscript}
\usetikzlibrary{calc,arrows,matrix,positioning, fit, trees}
\usepackage{acronym}
\usepackage{siunitx} 

\usepackage{balance}

\usepackage{makecell}

\usepackage{booktabs}
\usepackage{multirow}
\usepackage{rotating,tabularx}

\usepackage{algorithm,algorithmicx,algpseudocode,graphicx}
\providecommand{\keywords}[1]{\textbf{\textit{Keywords:}} #1}

\begin{document}
\title{Automatic Software and Computing Hardware Co-design for Predictive Control}
\author{Bulat Khusainov$^*$, Eric C.\ Kerrigan$^*{^\dagger}$,  George A.\ Constantinides$^*$%
\thanks{This work was funded from the People Programme (Marie Curie Actions) of the European Union's Seventh Framework Programme (FP7/2007-2013) under REA grant agreement no 607957 (TEMPO).}%
\thanks{B.\ Khusainov and G.\ A.\ Constantinides are with the Department of Electrical \& Electronics Engineering, Imperial College London, SW7~2AZ, U.K. {\tt\small b.khusainov@imperial.ac.uk}, {\tt\small g.constantinides@imperial.ac.uk}}%
\thanks{E.\ C.\ Kerrigan is with the  Department of Electrical \& Electronic Engineering
and Department of Aeronautics, Imperial College London, London SW7~2AZ, U.K. {\tt\small e.kerrigan@imperial.ac.uk}}
}
\maketitle

\begin{abstract}
Model Predictive Control (MPC) is a computationally demanding control technique that allows dealing with multiple-input and multiple-output systems, while handling constraints in a systematic way. The necessity of solving an optimization problem at every sampling instant often (i) limits the application scope to slow dynamical systems and/or (ii) results in expensive computational hardware implementations. Traditional MPC design is based on manual tuning of software and computational hardware design parameters, which leads to suboptimal implementations. This paper proposes a framework for automating the MPC software and computational hardware co-design, while achieving the optimal trade-off between computational resource usage and controller performance. The proposed approach is based on using a multi-objective optimization algorithm, namely BiMADS. Two test studies are considered: Central Processing Unit (CPU) and Field-Programmable Gate Array (FPGA) implementations of fast gradient-based MPC. Numerical experiments show that optimization-based design outperforms Latin Hypercube Sampling (LHS), a statistical sampling-based design exploration technique.

\end{abstract}

\section{Introduction and related work} \label{sec:co_design_intro}

Model predictive controller design is a multidisciplinary problem that involves tuning several coupled design parameters. Traditionally MPC controllers were tuned manually, with a trial and error approach, which cannot be considered as a viable option for most present-day applications, considering the number of design parameters and design evaluation time~\cite{simulation_verification}. Moreover, manual tuning often requires understanding the nature of the controlled dynamical system and MPC controller with the underlying optimization solver. Available tuning guidelines for model predictive control, including heuristic and systematic (but not automatic) approaches, are reviewed in~\cite{mpc_tuning}. Note that only high level optimal control problem parameters (e.g.\ horizon length, weights on states/inputs) are considered in the review paper, without regard to solving the underlying optimization problem.

The full design exploration approach, which can be considered as the simplest way of design automation, leads to unacceptable exponential complexity scaling with respect to the number of parameters.  Statistical exploration methods, e.g.\ Monte Carlo methods~\cite{monte_carlo_doe} and Latin Hypercube Sampling (LHS)~\cite{lhs} attempt to accelerate exploration by randomising the sampling process. However, all above techniques \textit{explore} the design space without \textit{exploiting} the knowledge about evaluated designs. As a result, statistical algorithms often achieve uniform distribution in the parameters space, without giving priority to the most promising (in terms of performance criteria) areas and hence waste time evaluating unpromising implementations. However, Monte Carlo methods and LHS can be used for identifying an initial guess for other algorithms.

An alternative approach for taking humans out of the design loop is based on systematic optimization~\cite{human_out_of_loop}. The following applications of optimization techniques to MPC design can be found in the literature:
\begin{itemize}
	\item \cite{tuning_mpc_vallerio} presents an application of multi-objective optimization to the Van de Vusse reaction considering two contradicting objectives: maximizing the desired product and minimizing the unwanted product. Several multi-objective optimization methods are employed in this work: normalized normal constraint~\cite{moo_nnc}, normal boundary intersection~\cite{moo_nbi} and weighted sum.
	\item Study~\cite{nmpc_tuning_biochemical} compares systematic multi-objective optimization-based parameter tuning using enhanced normalized normal constraint method~\cite{moo_enhanced_nc} against Monte Carlo simulations. The case study considers NMPC applied to a biochemical system.  
	\item Another algorithm for tuning NMPC controllers with application to chemical processes is presented in~\cite{nmpc_tuning_utopia}. Instead of exploring the trade-offs the algorithm attempts to find a single compromise design using the approach~\cite{zavala_utopia}. 
\end{itemize}
Other examples of automatic tuning of MPC controllers can be found in~\cite{Yamashita_tuning_mpc} and~\cite{Yamashita_tuning_mpc2}. The studies outlined above share certain common features:
\begin{itemize}
	\item Multi-objective nature of MPC design problem is acknowledged and hence dedicated multi-objective optimization algorithms are used.
	\item Contradicting control objectives are considered as design goals without explicit optimization of computational complexity. 
	\item Optimization solver parameters are not systematically tuned, it is assumed that optimal control problems are solved to optimality. This might be a valid assumption for slow dynamical systems, e.g.\ chemical reactors, but not for fast applications, e.g.\ robotics.
	\item Underlying computational platform parameters are not involved in the tuning process.
\end{itemize}

In contrast, in this work we automate MPC controller design considering computational resources as a design objective. The control performance can be traded off against resource usage by tuning both algorithmic and computational hardware-related parameters. 

The paper is organised as follows. Following the introduction, the target computational platform is described in Section~\ref{sec:fpga}. An approach for formulating predictive controller design as an optimization problem is presented in Section~\ref{sec:design_optimization_problem}. Possible ways of formalising design objectives and constraints are discussed within the section. Following that, Section~\ref{sec:moo} reviews existing algorithms for solving the resulting optimization problem and justifies using the BiMADS algorithm~\cite{bimads} for solving MPC design problems. Two case studies are considered in Section~\ref{chapter5_case_studies}: design of CPU-based and FPGA-based implementations of a fast gradient-based predictive controller. Section~\ref{chapter5_summary} concludes the paper.

\section{Target computational platform} \label{sec:fpga}

In this work we prototype predictive controller implementations using Xilinx Zynq-7000 XC7Z020 System-on-a-Chip (SoC), which incorporates a dual-core ARM Cortex-A9 processor and Artix-7 FPGA logic.
\begin{figure}[tb]
	\centering
	\includegraphics[width=0.75\columnwidth]{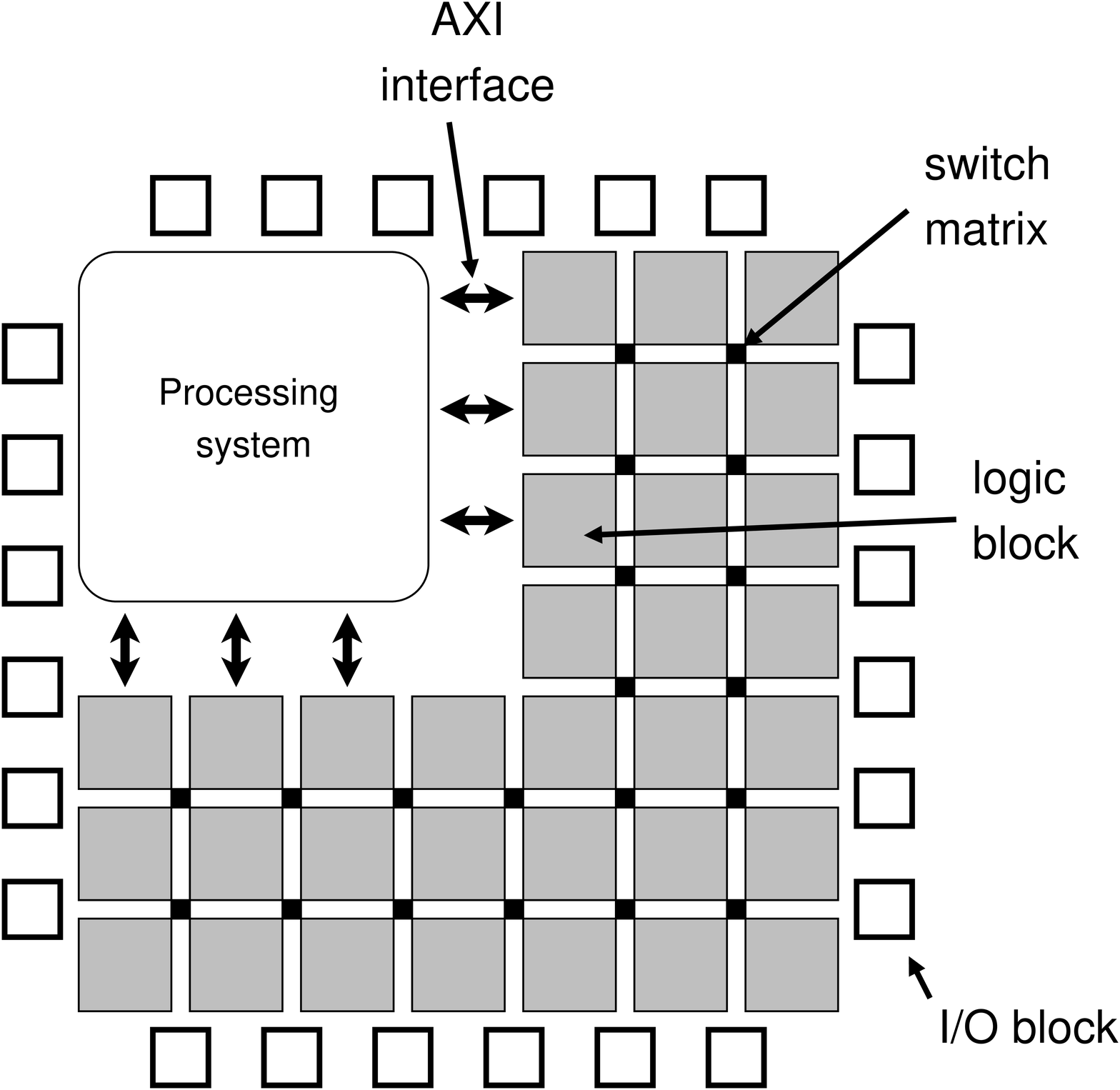} 
	\caption{Zynq SoC architecture.}
	\label{fig:zynq_arch}
\end{figure}
The FPGA fabric contains 53200 Lookup Tables (LUTs), 106400 Flip-Flops (FFs), 220 DSP blocks and 140 block RAMs (BRAMs) with total capacity of 4.9 Mb. Using SoC allows prototyping both Central Processing Unit (CPU) and FPGA implementations using one single chip. 

The following techniques can be used to accelerate optimization-based controllers on FPGAs:
\begin{itemize}
	\item \textit{Data-level parallelization} refers to splitting computations across multiple processors, so that different
sets of data are processed simultaneously. Usually applied on regular data structures, e.g. arrays.
	\item \textit{Data pipelining} is based on connecting processing units in series, so that the output of one unit is the input to the next one. The elements of this chain, i.e. pipeline, process data simultaneously.
	\item Tailoring \textit{number representation} for a given algorithm, instead of using standard single/double precision floating point arithmetic, allows achieving better time vs resource usage trade-offs. For example, for fixed-point number representation arithmetic operations complexity is identical to that of integers. 
\end{itemize}

In this work we synthesize FPGA circuits using vendor's high level synthesis tool, namely Vivado HLS. Vivado HLS allows describing FPGA circuit architecture with C code and additional optimization directives for implementing the above above-discussed acceleration techniques. The entire FPGA design flow involves multiple stages:
\begin{itemize}
	\item \textit{High-level synthesis}: Converting the C code with synthesis directives (e.g. loop unrolling or
pipelining) into low level Hardware Description Language (HDL) code.
	\item \textit{Synthesis}: The process of transforming HDL code into a netlist, a graph that defines the
connection of all logic gates and registers.
	\item \textit{Place-and-route (P}{\&}\textit{R)}: Solving a set of optimization problems in order to fit the netlist to a
particular FPGA platform. The outcome of P\&R is a bitstream that can be uploaded onto the
FPGA.
	\item \textit{Functional verification} of the circuit. For optimization-based control applications, the FPGA
circuit has to be verified in the loop with a plant model.
\end{itemize}  
In this work we automate the above design flow by using Protoip software tool~\cite{nmpc_soc}. 

\section{Formulating the design problem as an optimization problem} \label{sec:design_optimization_problem}


\subsection{Design objectives and constraints} \label{sec:design_objectives}

\subsubsection{Performance}

Quantifying controller performance is not a trivial task: depending on the nature of the dynamic system and design requirements several performance criteria might be considered. Fortunately, optimization-based controllers often perform explicit performance optimization and hence can be evaluated using objective measures. Depending on the application different control objectives might be selected, e.g. energy consumption or settling time. It should be emphasized that the solution of a single optimal control problem cannot serve as a performance indicator, since the ultimate goal is achieving desired closed-loop behaviour~\cite{eric_codesign}. Instead, a closed-loop cost function should be calculated based on a closed-loop simulation.

Considering the closed-loop cost function as an objective within an optimization framework, it is important to take into consideration the continuity and monotonicity properties of the cost function. According to~\cite{mpc_value_article}, even for a constrained LQR formulation, which is arguably the simplest MPC setup, neither continuity nor monotonicity with respect to horizon length and sampling time can be guaranteed in general. This significantly limits the range of optimization tools that can be used for MPC design optimization.

\subsubsection{Computation time} \label{sec:computation_time}

In relation to CPU implementations, where several algorithms might share the same hardware platform, algorithm execution time becomes both a design objective and design constraint. On the one hand, minimizing algorithm execution time  keeps processor load  low  and hence enables sharing processor time with other algorithms. On the other hand there is a fundamental constraint for MPC design problems: in order to implement the controller in real-time, the optimization algorithm execution time has to be smaller than the sampling time of the system. There are certain exceptions to this rule~\cite{intra_delay_sampling}, which are not considered in this work.

In contrast, for FPGA implementations, where a circuit is synthesised for one particular algorithm, minimizing computation time would not give any benefits, since the logic cannot be reused by other algorithms. Moreover, for certain cases it could be worthwhile to increase computation time by decreasing the circuit size and hence saving resources. However, computation time can only be increased up to the sampling time, which can be formalized as a design constraint. 

Note that, for a given algorithm the CPU implementation execution time would not be fully predictable because of a complex memory hierarchy. For FPGA circuits, execution time (in terms of clock cycles) can often be determined based on the architecture and hence efficiently predicted before circuit synthesis.



\subsubsection{Computational logic usage and energy consumption}

An FPGA designer often aims to minimize the amount of silicon that is required for implementing a particular control algorithm to get a size- and cost-efficient solution. As discussed in Section~\ref{sec:fpga}, a modern FPGA has the following resources: flip-flops, lookup tables, block RAMs and DSPs. We measure the silicon usage (or computational logic usage) as follows:

\begin{equation} \label{eq:fpga_resource_measure}
R_{FPGA} = \sqrt{R_{FF}^2 + R_{LUT}^2 + R_{DSP}^2 + R_{BRAM}^2},
\end{equation}
\noindent where $R_{FF}$, $R_{LUT}$, $R_{DSP}$ and $R_{BRAM}$ denote relative utilization of each resource type. Euclidean norm can be considered as a compromise between the $L^1$ and $L^{\infty}$ norms, where the former would not take into account a possible imbalance between different types of resources and the latter would penalize only the most heavily used resource.

 In contrast to taking the average (which in this case would be equivalent to the $L^1$ norm), the Euclidean norm automatically accounts for the fact that a certain implementation is constrained by the most heavily-used resource.

There is seldom a linear correspondence between circuit size and energy consumption. For instance, in some cases it could be energy-efficient to create a large circuit by parallelizing all computations in order to quickly  perform all calculations and switch to idle mode~\cite{eric_codesign}. In such cases, energy consumption may be considered as a separate design objective, which is particularly important for energy-autonomous embedded platforms.

\subsection{Design parameters} \label{sec:design_parameters}

\subsubsection{Horizon length and sampling time}

Horizon length and sampling time are fundamental design variables that have a crucial impact both on closed-loop performance of a system and computational hardware requirements. These two parameters, being tightly coupled with each other, define the quantity of predicted information in a predictive controller. Horizon length defines the ``vision distance'', while sampling rate sets the ``quality of the picture''~\cite{mpc_value_article}. Sampling frequency also defines the response delay of the controller. An overview of techniques for manual tuning of these parameters can be found in~\cite{mpc_tuning}.

\subsubsection{Problem formulation: condensed vs non-condensed} \label{sec:sparse_condensed}

Eliminating the states from decision variables by expressing them via the current state and input sequence leads to a compact condensed formulation~\cite{condensed_vs_sparse}, which results in a worst-case cubic scaling of the number of floating point operations in horizon length for the primal-dual interior point algorithm iteration. A sparse formulation implies the opposite approach: keeping the dynamics in constraints and treating both inputs and states as decision variables. Although the problem size becomes larger, exploiting the sparsity pattern allows linear scaling of computational effort in horizon length. The condensed and non-condensed formulations can be considered as two extreme points of a \textit{sparsity level} design variable. Controlling the level of sparsity can be achieved by dividing the prediction horizon into sub-intervals and performing partial condensing. This provides the possibility of adjusting the block size of linear algebra problems in order to find the optimal level of sparsity in terms software and hardware resources utilization~\cite{sparsity_level}. 

\subsubsection{Optimization algorithm parameters}

Optimization algorithms that solve optimal control problems numerically often have many tuning parameters on the underlying levels. The first fundamental design choice is the algorithm type~\cite{eric_codesign}: 
\begin{itemize}
	\item first vs second order methods
	\item interior point vs active set algorithms
	\item gradient-based vs derivative-free approaches
\end{itemize}

In addition to making high level design decisions, it is important to tune low level parameters, which might include:
\begin{itemize}
	\item number of iterations / termination criterion
	\item barrier parameter update strategy for interior-point algorithms
	\item globalization strategy (line search vs trust region)
\end{itemize}

The above parameters must be tuned with respect to closed-loop performance, which is not necessary correlated with a single optimization problem's optimality conditions~\cite{eric_codesign}. For example, a sub-optimal controller with a longer prediction horizon might perform better than an optimal controller with a shorter horizon.  More details on sub-optimal MPC can be found in~\cite{suboptimal_mpc}. 

\subsubsection{Number representation} \label{subsec:data_rep}

There are usually two types of choices that have to be made in relation to number representation:

\begin{itemize}
	\item The conceptual choice of data representation type: e.g.\ floating point or fixed point.
	\item The numbers of bits to be allocated for different parts of a number, e.g. for mantissa and exponent in floating point arithmetic.
\end{itemize}

\subsubsection{Data-level parallelism and pipelining} \label{subsec:parallelization}

Data-level parallelism and pipelining were discussed in Section~\ref{sec:fpga}. The main algorithmic choices in relation to these techniques are:

\begin{itemize}
	\item The number of parallel processors.
	\item The number of pipeline stages, i.e. \textit{pipeline depth}.
\end{itemize}

It should be emphasized that parallelism and pipelining affect algorithm execution time and resource usage, which may or may not have an impact on the closed-loop performance.

\subsection{The resulting optimization problem}

The design parameters considered in Section~\ref{sec:design_parameters} can be classified in two categories: software parameters (e.g. prediction horizon length) and hardware parameters (e.g. number representation). Conventional approaches propose sequential design: initially the software algorithm is designed at a high level of abstraction without regard to the intended hardware platform and, following this, the algorithm is implemented on a hardware platform by selecting appropriate hardware parameters. This decoupled approach usually leads to inefficient utilization of available resources~\cite{eric_codesign}. In contrast, the co-design approach implies simultaneous design of both software and hardware components in order to achieve the best possible performance for a given set of available resources. However, improvement of the closed-loop performance cannot be considered as the only design objective. As can be seen from Section~\ref{sec:design_objectives} there are often several contradicting design objectives, which might include performance and computational hardware resource usage. Instead of looking for one optimal design (which often does not exist due to conflicts between objectives), engineers might make a decision based on the whole series of Pareto optimal designs, i.e.\ designs that cannot be improved in terms of one objective without worsening at least one of the other objectives.

The problem of investigating design trade-offs is usually formalized as a multi-objective optimization (MOO) problem. The main bottleneck that prevents efficient solution of MOO design problems in relation to MPC are the properties of the objective functions:
\begin{itemize}
\item Long function evaluation time. Evaluation of the design objective functions requires time-consuming simulations to be performed.
\item Absence of derivative information. There are no analytical expressions for accurate estimation of the design objectives derivatives.
\item Mixed domain.  Design variables can be both discrete (e.g.\ number of bits for data representation) and continuous (e.g.\ sampling rate).
\item Noisiness. For example, the same HLS code may result in different resource usage values depending on a vendor's software version or other factors that are not taken into account by conventional models.
\end{itemize}

The next section will review existing algorithms for solving multi-objective optimization problems with focus on BiMADs, a bi-objective version of a mesh adaptive direct search algorithm.

\section{Derivative-free multi-objective optimization} \label{sec:moo}

\subsection{Problem statement}

We consider the following multi-objective optimization problem:
\begin{subequations} \label{eq:mo_codesign}
		\begin{align}
		\underset{p}{\textrm{min}} &  \quad f(p) \coloneqq \begin{pmatrix}  f^{(1)}(p) \\ \vdots \\ f^{(l)}(p)  \end{pmatrix} \\
			 \textrm{subject to}  &  
			 \quad p \in S,  
		\end{align}	\end{subequations}
\noindent
where $S$ is $q$-dimensional decision space. Since the objectives of MOO are often contradicting, there is no single solution to the problem. Instead, a set of \textit{Pareto optimal} solutions can be obtained.

\theoremstyle{definition}
\begin{definition}{Pareto optimality.}
A feasible solution $p^* \in S$ is Pareto optimal if there does not exist another feasible solution $p \in S$ such that $f_i(p) \leq f_i(p^*)$ for all $i \in \{1,\dots ,l\}$ and $f_i(p) < f_i(p^*)$ for at least one index $j \in \{1,\dots ,l\}$.
\end{definition}
The Pareto frontier is the set of all existing Pareto optimal points.

\theoremstyle{definition}
\begin{definition}{Pareto dominance.} \label{def:dominance}
A point $y' = f(p')$ is said to (strongly) dominate $y'' = f(p'')$ iff $\forall i \in \{1,\dots,l\}: y'_i  \leq y_i '' \textrm{ and } y' \neq y''$. The shorthand notation for this is $y' \prec y''$. Analogously, for weak dominance, $y' \preceq y''$ means $\forall i \in \{1,\dots,l\}: y'_i  \leq y''_i$.
\end{definition}

$P(U)$ is the set of non-dominated points for a given set of evaluated points $U$, i.e.\ the current approximation of the Pareto frontier. The quality of a Pareto frontier approximation can be assessed by means of the \textit{hypervolume}.

\theoremstyle{definition}
\begin{definition}{Hypervolume.} \label{def:hypervolume}
For a given reference point $y_{ref}$ and Pareto frontier approximation~$P$, the hypervolume is defined as a set of points in the objective space $\{y \preceq y^{ref} \in \mathbb{R}^{l} | \exists y' \in P : y' \preceq y  \}$.
\end{definition}

The Lebesgue measure of the hypervolume $L(P, y_{ref})$, i.e. hypervolume space, defines the quality of the Pareto frontier approximation.

\subsection{Review of algorithms for derivative-free multi-objective optimization}

\subsubsection{Derivative-free single-objective optimization}

Deterministic algorithms for single-objective derivative-free optimization can be classified into~\cite{intro_derivative_free}:

\begin{itemize}
\item \textit{Trust-region interpolation algorithms}. Trust-region algorithms propose building a local approximation of the objective function based on evaluated samples. Based on this approximation, the function is minimized inside the trust region. 
\item \textit{Line search algorithms} for derivative-free optimization are conceptually similar to their derivative-based counterparts: they perform search along a particular direction. However, for derivative-free algorithms, the search direction is calculated without gradient information. 
\item \textit{Direct Search Methods} (DSMs) do not attempt to approximate derivatives either explicitly or implicitly. Instead, optimization is based on evaluation of a finite set of points around the current solution guess, so that a point with smaller objective function value can be found. 
\end{itemize}

Among the considered classes of algorithms, only direct search methods of directional type have been extensively studied in relation to multi-objective optimization~\cite{moo_derivative_free}, hence rest of the section will be focused on DSMs. 

Each iteration of a DSM is split into two parts: a \textit{search step} and \textit{poll step}. The latter step must be rigidly defined to guarantee convergence, while the former is more flexible and can be tuned to improve numerical performance. Assuming there is a given feasible solution guess, the general structure of a DSM of directional type can be described as follows~\cite{intro_derivative_free}:

\begin{enumerate}
	\item \textbf{Search step}. Evaluate the objective function at a finite set of points. If any of these points provide a better objective value compared to the current guess, declare the iteration as successful and skip the poll step.
	\item \textbf{Poll step}. Perform a local search around the current best point by evaluating a set of poll points, which are defined by poll directions and a step size. Depending on the result of this search, declare the iteration as successful or unsuccessful.
	\item \textbf{Mesh parameter update}. Reduce step size for unsuccessful iterations, increase or maintain step size for successful ones. 
\end{enumerate}

Practical examples of DSM algorithms of directional type are the coordinate-search method~\cite{intro_derivative_free} and the Mesh Adaptive Direct Search (MADS) algorithm~\cite{mads}.

\subsubsection{Derivative-free multi-objective optimization}

There are two fundamentally different ways for tackling multi-objective optimization problems:

\begin{itemize}
	\item \textit{Direct} algorithms for multi-objective optimization attempt to approximate the Pareto frontier directly.
	\item \textit{Scalarization-based} approaches propose converting the multi-objective problem into a sequence of single-objective problems. 
\end{itemize}

The Direct Multisearch (DMS) algorithm~\cite{dms_algorithm}, which belongs to the class of direct multi-objective optimization algorithms, is a generalization of DSM algorithms for single-objective optimization. The main components of a DMS are the same as DSM as discussed in the previous subsection: the search step, the poll step and mesh update. The algorithm keeps a list of non-dominated points, which  represents the current Pareto frontier approximation. The local poll search is performed around several non-dominated points. Successfulness of an iteration is decided based on the changes in the Pareto frontier approximation. The search step, similarly to single-objective DSM algorithms, is flexible and is not required for convergence~\cite{moo_derivative_free}. However, it can help to improve the distribution of the points along the Pareto frontier, although this is not a systematic way of ensuring uniformity of the frontier.

A classical example of scalarization-based algorithms is the weighted-sum method. The method scalarizes the vector of objectives into a single objective by taking an affine combination of all the objectives. Tuning the weights of the scalarized function allows one to move along the Pareto frontier. However, in this case movement along the frontier happens in a non-systematic way. As a result, some regions become over-represented and others may suffer from lack of information. In addition, this approach cannot deal with non-convexities and discontinuities in the Pareto frontier and hence loses some Pareto optimal solutions. 

The BiMADS algorithm~\cite{bimads} performs scalarization in a different way. For a problem with two objectives the scalarized function is given by
\begin{equation} \label{eq:bimads_formulation}
	\phi _r (p) = \left\{
                \begin{array}{ll}
                   - \prod\limits_{i=1}^{2} (r_i - f^{(i)}(p))^2  & \textrm{if } f(p) \preceq r, \\
                     \sum\limits_{i=1}^{2} ( \textrm{max} \{0,(f^{(i)}(p) - r_i) \} )^2 & \textrm{otherwise,}
                \end{array}
              \right.
\end{equation}
where $r$ is a reference point in the objective space. The problem of minimizing $\phi _r(\cdot)$ is solved by the MADS~\cite{mads} algorithm. The formulation~\eqref{eq:bimads_formulation} with appropriate selection of a reference point allows the recovery of all Pareto optimal solutions, which is not the case for the weighted-sum reformulation. For achieving good distribution of non-dominated solutions, the reference point must be selected in a particular way. The BiMADS algorithm proposes a formal method for identifying the biggest gap in the Pareto frontier approximation and presents a procedure for selecting an appropriate reference point. Details can be found in~\cite{bimads}. 

In this work, the BiMADS algorithm will be used for solving the multi-objective co-design problem. The main advantages of the algorithm are:

\begin{itemize}
	\item Both BiMADS and the underlying single-objective MADS solver are supplied with a mathematical proof of convergence~\cite{bimads, mads}, which is not the case for other considered algorithms.
	\item BiMADS has proven to be efficient for solving real-world engineering problems~\cite{bimads_material_science}.
	\item There is a free software application that implements the algorithm, namely NOMAD~\cite{nomad}
	\item NOMAD provides various interfaces, including a Matlab front-end, which simplifies integration with Protoip~\cite{nmpc_soc}.
\end{itemize}

The application scope of BiMADS is limited to bi-objective problems, which can be considered as the main drawback of the algorithm.


\section{Case studies} \label{chapter5_case_studies}

\subsection{Fast gradient-based controller for a mass-spring-damper system - CPU-only implementation} \label{sec:chapter5_case1}

\begin{figure}[t]
	\centering
	\includegraphics[width=0.8\columnwidth]{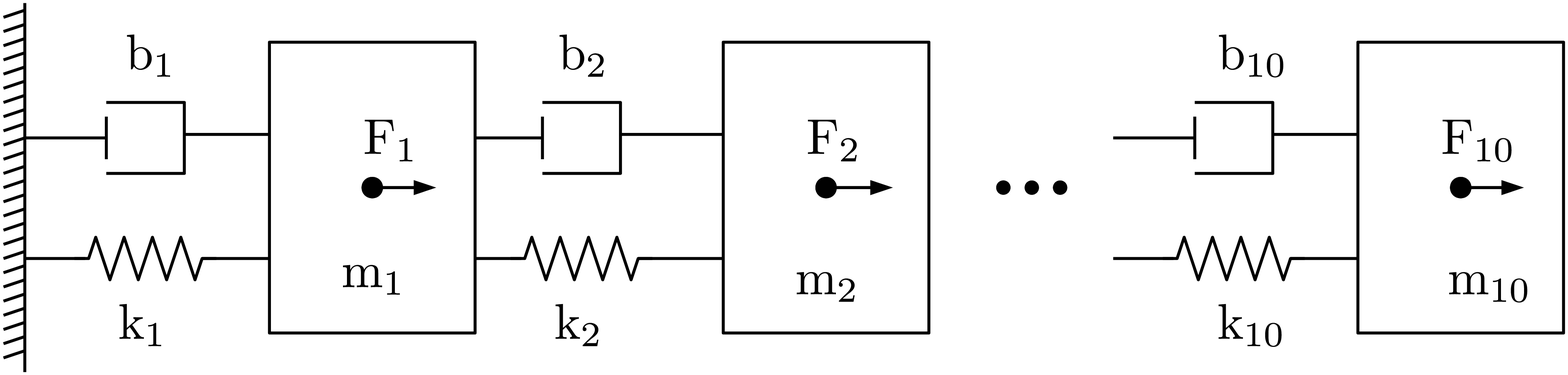} 
	\caption{Mass-spring-damper system.}
	\label{fig:mass_spring}
\end{figure}

The system under control represents a chain of ten masses that are connected via springs and dampers (Figure~\ref{fig:mass_spring}). 
The first mass is also connected to a fixed wall. Each mass can be actuated with an input force that has input and output limits. It is assumed there is no gravitational force. The system  can be modelled with a continuous-time linear state space model:
\begin{equation}
	\label{cont_linear_ss_model}
	\dot{x}(t) = A_{c}x(t) +B_{c}u(t)		
\end{equation}
\noindent	 
where $A_{c} \in \mathbb{R}^{n \times n}$, $B_{c} \in \mathbb{R}^{n \times m}$ are continuous-time state and input matrices. For a system of ten masses $n=20$ and $m=10$.

The following optimal control formulation is considered:
\begin{subequations} \label{cont_linear_ocp}
\begin{align} 
		\underset{u,x}{\textrm{minimize}} &  \quad \int_0^T \left( \frac{1}{2}x^{T}(t) Q_{c}x(t) + \frac{1}{2}u^{T}(t) R_{c}u(t) \right. \nonumber
		\\
		& \left. + x^{T}(t) W_{c}u(t) \right)  dt + \frac{1}{2}x^{T}(T)P_c x(T)  \\
			 \textrm{subject to }  &  
			  x(0) = \hat{x}   \\
			 &  \dot{x}(t) = A_{c}x(t) +B_{c}u(t), \ \forall t \in \lbrack 0,T \rbrack    \label{cont_linear_dynamics}  \\
			 & \quad u_{min} \leq u(t) \leq u_{max}, \ \forall t \in \lbrack 0,T \rbrack\end{align}
\end{subequations}
where $Q_{c} \in \mathbb{S}_{+}^{n}$, $R_{c} \in \mathbb{S}_{++}^{m}$, $W_{c} \in \mathbb{R}_{}^{n \times m}$ and $P_d \in \mathbb{S}_{++}^{n}$ are state, input, cross and terminal penalty matrices accordingly. $\mathbb{S}_{++}^{n} (\mathbb{S}_{+}^{n})$ denotes a set of positive (semi-)definite matrices. 

In this experiment the following prediction matrices were used:
\begin{multline} \label{eq:prediction_matrices}
R_c = I_{m \times m} \otimes [0.0001], \quad Q_c = I_{n \times n} \otimes \begin{bmatrix} 1 & 0 \\ 0 & q_{speed}   \end{bmatrix}, \\
W_c = 0_{m \times n}, \quad P_c = Q_c,
\end{multline}
where $I$ and $0_{m \times n} $ denote identity and zeros matrices accordingly. Tuning $q_{speed}$, which is a design parameter, allows the changing the ratio between penalising masses positions and velocities.

For the purpose of digital control, the continuous-time state-space model~\eqref{cont_linear_ss_model} and optimal control problem~\eqref{cont_linear_ocp} are discretized, assuming a zero-order hold:
\begin{subequations} \label{discr_linear_ocp}
\begin{align} 
		\underset{u_0 \dots u_{N-1}, x_0 \dots x_N}{\textrm{minimize}} &  \quad \sum_{k=0}^{N-1} \left( \frac{1}{2}x_{k}^{T} Q_{d}x_{k} + \frac{1}{2}u_{k}^{T} R_{d}u_{k} \right. \nonumber\\
		& \left. + x_{k}^{T} W_{d}u_{k} \right) + \frac{1}{2}x_{N}^{T} P_d x_{N} \\
			 \textrm{subject to }  &   x_{0} = \hat{x}   \\
			  x_{k+1} = A_d x_{k} + &B_d u_{k},  \ \forall  k \in\{0,\ldots,N-1\}  \label{linear_dynamics}  \\
			  u_{min} \leq u_{k} \leq &u_{max},   \ \forall  k \in\{0,\ldots,N-1\}
			 \end{align}
\end{subequations}
\noindent
Note that the discrete-time penalty matrices  $Q_d$, $R_d$, $W_d$ and $P_d$ are model-dependent~\cite{mpc_value_article}.

The optimal control problem~\eqref{discr_linear_ocp} can be transformed into a quadratic programming problem by eliminating the states, which leads to the condensed formulation
\begin{subequations}	\label{eq:QP}
		\begin{align}
		\underset{}{\textrm{minimize}} & \quad  \frac{1}{2}\theta^{T}H\theta + \theta^{T}h \label{eq:qp_objective} \\
			 \textrm{subject to}  &  
			 \quad \theta_{min} \leq \theta \leq \theta_{max}  \label{eq:input_constr}
		\end{align}
	\end{subequations}
Note that the gradient term $h$ depends on the current state, while the Hessian $H$ is fixed and hence can be precalculated offline. More details on condensed and sparse formulations for predictive control can be found in~\cite{condensed_vs_sparse}. Since~\eqref{eq:input_constr} has the form of box constraints, calculating projection on the feasible set becomes computationally cheap.  This facilitates using Nesterov's projected gradient algorithm~\cite{nesterov_fgm}, also known as the Fast Gradient Method (FGM). The method proposes moving in the anti-gradient direction  and performing projection $P(\cdot)$ on the feasible set after each iteration (Algorithm~\ref{fgm_algorithm}). The extra momentum step with a parameter $\beta$ achieves an optimal convergence rate. The constant step scheme~\cite{nesterov_fgm} implies $\beta = \frac{\sqrt{L} - \sqrt{\mu}}{\sqrt{L} + \sqrt{\mu}}$, where $L$ is the largest eigenvalue of the Hessian $H$ and $\mu$ is the convexity parameter, which is equal to minimum eigenvalue of the Hessian.
\begin{algorithm}
		\caption{Projected fast gradient algorithm for constrained optimization with constant step size.}
		\label{fgm_algorithm}
		\begin{algorithmic}[1]
		\State $\textrm{Initial guess:}\theta_0$
		\State $ v_0 = \theta_0$
		\For{$i=0$ to $N_{FGM}$}
		\State $\theta_{i+1} = (I-(1/L)H) \nu_{i} - (1/L) h$ \Comment{anti-gradient step}
		\State $z_{i+1} = P(\theta_{i+1})$ \Comment{projection on the feasible set} 
		\State $v_{i+1}=(1+\beta)z_{i+1} - \beta z_i $ \Comment{extra-momentum step} 
		\EndFor
		\end{algorithmic}
\end{algorithm}

\begin{figure}
	\centering
	\includegraphics[width=0.8\columnwidth]{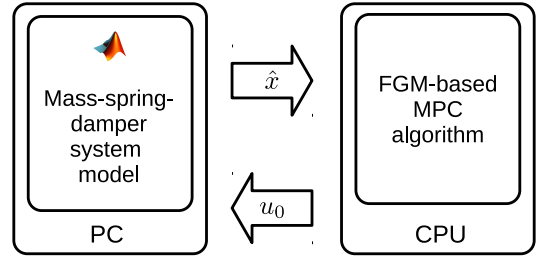} 
	\caption{CPU in the loop test setup.}
	\label{fig:cpu_in_the_loop}
\end{figure}

The algorithm was implemented on an ARM Cortex A9 processor of the Xilinx Zynq-7000 XC7Z020 system-on-a-chip employing only a single processing core. Using the Protoip toolbox allowed for fast verification of the controller in the loop with the plant model as shown in Figure~\ref{fig:cpu_in_the_loop}. 
 
The problem of interest is automatic design of a fast gradient-based controller. The following design objectives are considered:
\begin{itemize}
	\item \textit{Controller performance} is judged based on settling time. In this experiment, settling time is defined as the time elapsed from the beginning of closed-loop simulation to the time at which $\left\lVert x(t) \right\rVert _2 \leq \epsilon$, where $\epsilon = 0.01$. Several simulations with different initial conditions are performed in order to calculate performance measure as a sum of settling times for different initial conditions; the list of initial conditions are presented in Appendix~\ref{sec:appendix1}.  Since the performance criterion and MPC objective are different, it is essential to tune the prediction matrices in order to achieve the desired performance.
	\item \textit{Algorithm computational time}. As discussed in Section~\ref{sec:computation_time}, computational time is the main measure of algorithm complexity for CPU implementations.
\end{itemize}

Design constraints:
\begin{itemize}
	\item \textit{Algorithm computational time}, in addition to being a design objective, appears in a constraint function: in order to implement the controller in real-time the algorithm execution time has to be smaller than the sampling time of the system (Section~\ref{sec:computation_time}).
	\item \textit{Stability constraint} captures whether the controller was able to stabilize the system. Unstable response might happen due to short horizon, numerical errors or other reasons.    
\end{itemize}

Note that the former constraint is \textit{quantifiable} while the latter is \textit{non-quantifiable}. A quantifiable constraint is a constraint for which the degree of feasibility and violation can be quantified~\cite{progressive_barrier}. In this work, non-quantifiable constraints are handled with the \textit{extreme barrier} approach, which implies setting the objective to infinity for all infeasible points and therefore not allowing infeasible iterations. For quantifiable constraints the \textit{progressive barrier} approach is adopted. Progressive barrier constraint handling allows exploiting knowledge of the violation degree by accepting infeasible iterations. Both extreme and progressive barrier approaches are implemented in NOMAD. More details can be found in~\cite{progressive_barrier, nomad}.

The design parameters are the following:

\begin{itemize}
	\item Horizon length, $N$ in~\eqref{discr_linear_ocp}; bounds: $1 \leq N \leq 12. $
	\item Sampling time, $T_s$; bounds: $0.02 \leq T_s \leq 0.5. $
	\item Number of fast gradient algorithm iterations, $N_{FGM}$ in Algorithm~\ref{fgm_algorithm}; bounds: $20 \leq N_{FGM} \leq 200. $
	\item State penalty matrix, particularly $q_{speed}$ parameter in~\eqref{eq:prediction_matrices}; bounds:  $0.2 \leq q_{speed} \leq 5. $
\end{itemize}

\begin{figure}[t]
  \centering
	\begin{subfigure}[]{\columnwidth}
		\centering
			\includegraphics[width=0.7\columnwidth]{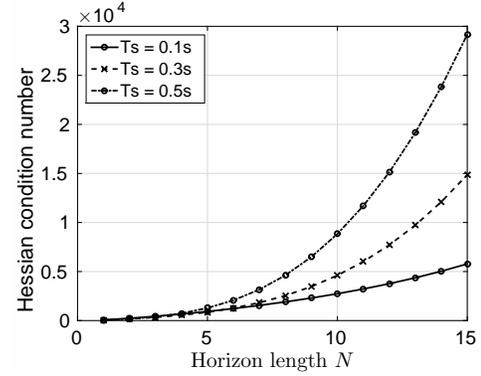} 
		\caption{The impact of horizon length.}
		\label{fig:cond_num_horizon_length}
	\end{subfigure}
	\begin{subfigure}[]{\columnwidth}
		\centering
		\includegraphics[width=0.6\columnwidth]{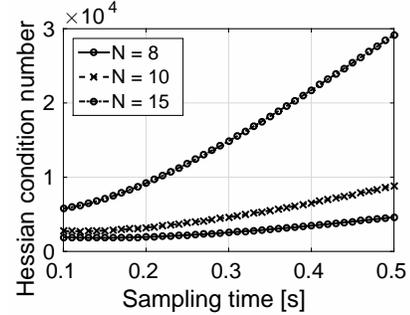} 
		\caption{The impact of sampling time.}
		\label{fig:cond_num_sampling_time}	
	\end{subfigure}
	\caption{The impact of horizon length and sampling time on Hessian condition number.}
	\label{fig:cond_num}
\end{figure}

The above parameters are tightly coupled with each other. For example, consider Figure~\ref{fig:cond_num} that illustrates the impact of horizon length and sampling time on the Hessian condition number. 
It can be observed that both parameters have a significant impact on the condition number, which in turn affects the convergence rate of the fast gradient algorithm~\cite{nesterov_fgm}. As a result, the number of fast gradient algorithm iterations $N_{FGM}$ required for convergence will also change. However, $N_{FGM}$ must be selected with respect to closed-loop performance, rather than open-loop optimality conditions, which complicates the tuning process even more.

\begin{figure}
\centering
\includegraphics[width=0.9\columnwidth]{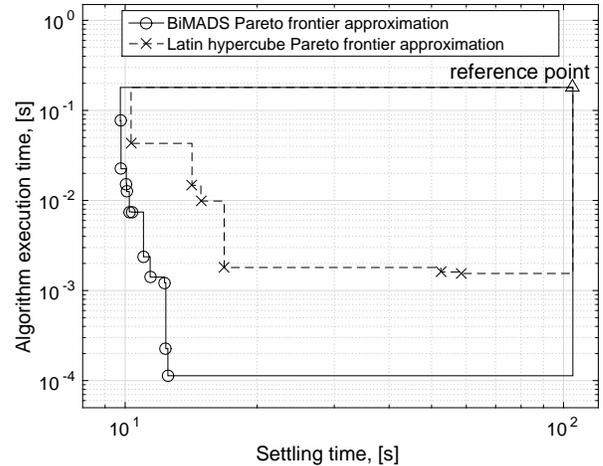} 
\caption{Pareto frontier approximation for CPU implementations of the FGM: BiMADS vs LHS.}
\label{fig:test_case1}	
\end{figure}

\begin{figure}
\centering
\includegraphics[width=0.9\columnwidth]{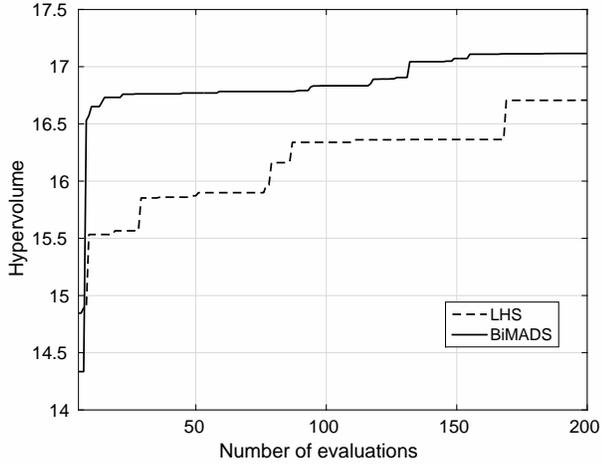} 
\caption{Hypervolume profiles for CPU implementations of the FGM: BiMADS vs LHS.}
\label{fig:test_case1_hv}	
\end{figure}

The above design problem was solved using the BiMADS algorithm. The results are compared to latin hypercube sampling, which is a statistical sampling method commonly used for design exploration and for design-of-experiments in particular~\cite{lhs}. For both experiments the number of evaluations was restricted to 200. Design evaluation involves compiling the source code and performing processor-in-the-loop tests, which takes 1-4 minutes for the considered setup, depending on a design complexity and the sampling time. Observe that full design exploration is not a viable approach: even using a coarse grid with ten points for the continuous variables, full exploration will require 217200 evaluations.  As can be seen from Figure~\ref{fig:test_case1},
the Pareto frontier returned by BiMADS dominates the front identified by LHS, i.e. any design sampled by LHS is dominated by at least one design identified by BiMADS (see Definition~\ref{def:dominance}). Moreover, according to the hypervolume profile (Figure~\ref{fig:test_case1_hv}), BiMADS provides a satisfactory approximation of the Pareto frontier on the early stages,
which opens the possibility of early termination, depending on the time and/or simulation resource availability.

Consider Table~\ref{tab:testcase1}, which lists designs identified by BiMADS. 
\begin{table}
\caption{Pareto frontier approximation identified by BiMADS for CPU implementations of FGM (corresponds to Figure~\ref{fig:test_case1}). Designs are sorted according to their settling times in ascending order.}
  \centering
  \begin{tabular}{c*{6}{c} }
    \hline \hline 
    	\multicolumn{4}{c}{Design parameters} &	\multicolumn{2}{c}{Design objectives}\\
    $T_s$, s & $N$ & $N_{FGM}$ & $q_{ratio}$ & \makecell{Settling \\ time, s } & \makecell{Algorithm \\ time, s} \\ \hline 
    0.1880  &  8  &  184 & 0.20 & 9.7499  & 0.0769 \\ 
    0.2360  &  5  &  136 & 0.20 & 9.7845  & 0.0225 \\
    0.3320  &  4  &  136 & 0.20 & 10.0598 & 0.0153 \\
    0.2360  &  7  &   40 & 0.20 & 10.0805 & 0.0126 \\
    0.3320  &  3  &  112 & 0.20 & 10.2197 & 0.0075 \\
    0.3800  &  3  &  112 & 0.20 & 10.3747 & 0.0074 \\
    0.2840  &  4  &   20 & 0.20 & 11.0092 & 0.0024 \\
    0.2840  &  2  &   40 & 0.20 & 11.4023 & 0.0014 \\
    0.4760  &  1  &  256 & 0.20 & 12.3020 & 0.0012 \\
    0.5000  &  1  &   44 & 0.20 & 12.3816 & 0.0002 \\
    0.4760  &  1  &   20 & 0.20 & 12.5158 & 0.0001 \\ \hline \hline \noalign{\vskip 0.03in}
  \end{tabular}
\label{tab:testcase1}
\end{table}
Designs are sorted according to their settling times in ascending order. It can be observed that most of the design parameters ($T_s$, $N$ and $N_{FGM}$) are changing along the Pareto-frontier in a non-monotonic way. This can be explained by the complicated cross-coupling between parameters. As a consequence, manual and heuristic tuning approaches are unlikely to identify these trade offs in an efficient way. The only exception is the weight matrix parameter $q_{ratio}$, which has the same value for all designs. In this particular case $q_{ratio}$ could be optimized separately from other design parameters.

Given the Pareto frontier (Figure~\ref{fig:test_case1}) a designer will be able to select a particular implementation based on the available processing time. For example:
\begin{itemize}
	\item For the computational budget of 100~ms, LHS-based exploration achieves a settling time of 10.31~s, while optimization-based design allows settling of the plant in 9.75~s.
	\item Tightening the computational budget to 2~ms leads to sampling times of 16.82~s and 11.40~s for LHS- and optimization-based approaches accordingly.
\end{itemize}

\subsection{Fast gradient-based controller for a mass-spring-damper system - FPGA implementation} \label{sec:chapter5_case2}

\begin{figure}
	\centering
	\includegraphics[width=0.8\columnwidth]{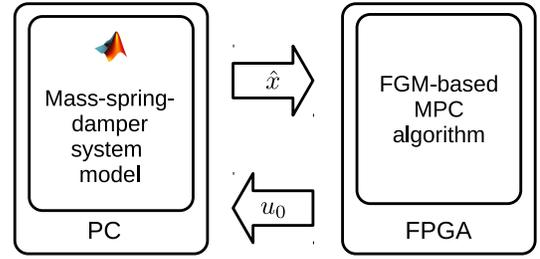} 
	\caption{FPGA in the loop test setup.}
	\label{fig:fpga_in_the_loop}
\end{figure}

This case study considers implementation of the Algorithm~\ref{fgm_algorithm} with fixed point arithmetic on the FPGA logic of Xilinx Zynq-7000 XC7Z020 SoC. The testing setup is similar to that of Section~\ref{sec:chapter5_case1}, both in terms of the optimal control problem formulation~\eqref{cont_linear_ocp} and the plant model (Figure~\ref{fig:mass_spring}). The algorithm was implemented with the Vivado HLS FPGA synthesis tool using Protoip for automatic deployment and verification in the loop with the plant model (Figure~\ref{fig:fpga_in_the_loop}).
As can be seen from Algorithm~\ref{fgm_algorithm}, FGM relies only on addition and multiplication operators, while all divisions can be precalculated offline. The following techniques were used to accelerate vector-vector and matrix-vector operations (lines 4-7):
\begin{itemize}
	\item \textit{Loop pipelining}~\cite{vivado_hls_guide}. Data pipelining, as a general acceleration technique, was discussed in Section~\ref{sec:fpga}. In relation to loops, pipelining implies overlapping iterations, i.e. starting a new iteration before finishing the previous. 
	\item \textit{Loop flattening}~\cite{vivado_hls_guide} is transforming nested loops into a single loop with multiple counters. Flattening allows efficient pipelining of nested loops. This technique was applied to matrix-vector multiplication (line 4), where loop nests arise when iterating over matrix rows and columns.
\end{itemize}

Fixed point arithmetic often introduces overflow and round off errors. The former issue can be addressed by precalculating the upper bound on the largest absolute value of algorithm iterates using interval arithmetic. Regarding overflow errors, the number of fraction bits has to be sufficiently large to maintain numerical stability of an iterative algorithm. A procedure for precalculating the number of integer and fraction bits for fixed point implementations of a fast-gradient algorithm is presented in~\cite{juan_megahertz}.

The following objectives are considered in this design optimization experiment: 
\begin{itemize}
	\item \textit{Controller performance} is measured similarly to the previous case study (Section~\ref{sec:chapter5_case1}) with $\epsilon = 0.02$.
	\item \textit{FPGA logic usage}. As discussed in Section~\ref{sec:design_objectives} FPGA designers often aim to minimize the amount of logic used for a particular algorithm. Logic usage is measured as the Euclidean norm of relative utilization of each resource type, see~\eqref{eq:fpga_resource_measure} in Section~\ref{sec:design_objectives}.
\end{itemize}

Design constraints:
\begin{itemize}
	\item \textit{Algorithm execution time}. Similarly to the previous case study, in order to implement the controller in real-time, the algorithm execution time has to be smaller than the sampling time of the system. This constraint is treated with the progressive barrier approach. Note that for FPGA setup (in contrast to CPU), computational time does not appear as an objective. This is explained by the fact that FPGA logic is synthesized for a particular algorithm and cannot be reused for other applications.
	\item \textit{Objective function convexity constraint}. Due to assumptions on the weight matrices in formulation~\eqref{cont_linear_ocp}, the Hessian of the objective function~\eqref{eq:qp_objective} is positive definite. However, a fixed point representation of the true Hessian may be non-convex because of truncation errors, which might affect convergence of the fast-gradient algorithm. To avoid non-convex formulations, a positivity constraint on the smallest eigenvalue of the fixed point representation of the Hessian must be set. Although this is a quantifiable constraint, which potentially can be treated with a progressive barrier approach, we will use the extreme barrier method that rejects all infeasible iterations. Since identifying Hessian convexity is significantly faster compared to the full design evaluation, which involves circuits synthesis and closed loop simulation, rejecting infeasible iterations allows saving design time.
\end{itemize}

The following design parameters are considered:
\begin{itemize}
	\item Horizon length, $N$ in~\eqref{discr_linear_ocp}; bounds: $1 \leq N \leq 12. $
	\item Sampling time, $T_s$; bounds: $0.02 \leq T_s \leq 0.5. $
	\item Number of fast gradient algorithm iterations, $N_{FGM}$ in Algorithm~\ref{fgm_algorithm}; bounds: $20 \leq N_{FGM} \leq 200. $
	\item State penalty matrix, particularly $q_{speed}$ parameter in~\eqref{eq:prediction_matrices}; bounds:  $0.2 \leq q_{speed} \leq 5. $
	\item Number of fraction bits for fixed point number representation, $N_{frac}$; bounds: $5 \leq N_{frac} \leq 25.$
\end{itemize}

Similarly to the previous case study, the multi-objective optimization problem with the above design objectives and constraints was solved using BiMADS algorithm and results were compared to Latin hypercube sampling allowing 200 evaluations for both algorithms. In contrast to software compilation, FPGA circuit synthesis is a time-consuming process, which leads to the design evaluation time in the range of 20-35 minutes. For this case study full design exploration would require 4561200, assuming a grid of ten points for the continuous parameters.

\begin{figure}
\centering
\includegraphics[width=0.9\columnwidth]{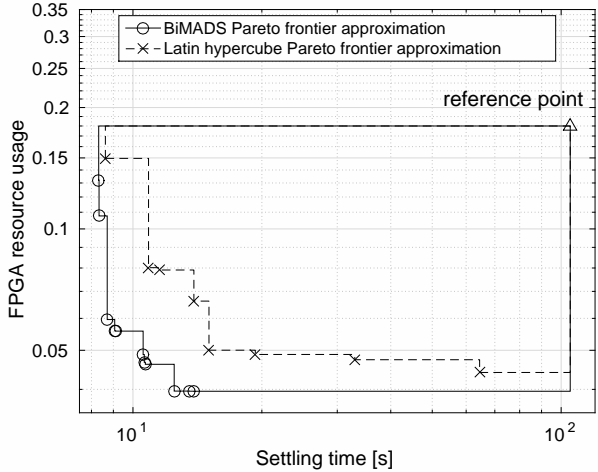} 
\caption{Pareto frontier approximation for FPGA implementations of the FGM: BiMADS vs LHS.}
\label{fig:test_case2}	
\end{figure}

\begin{figure}
\centering
\includegraphics[width=0.9\columnwidth]{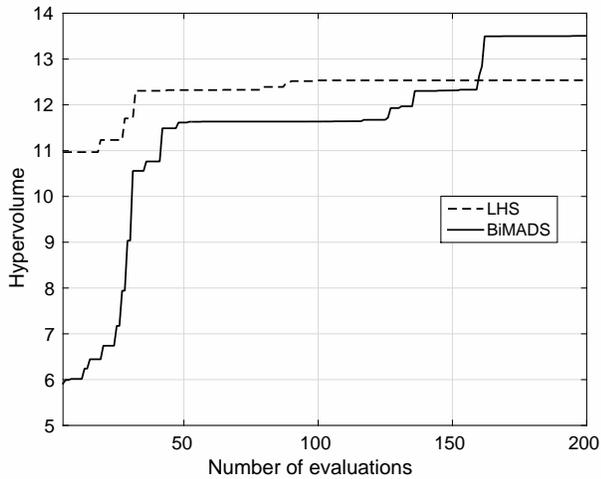} 
\caption{Hypervolume profiles for FPGA implementations of the FGM: BiMADS vs LHS.}
\label{fig:test_case2_hv}	
\end{figure}

It can be observed from Figure~\ref{fig:test_case2} that BiMADS outperforms LHS, i.e. the BiMADS Pareto frontier dominates frontier identified by LHS. 
However, unlike with the previous case study, LHS outperforms BiMADS in the early stages of design exploration, which can be visualized with the hypervolume profile in Figure~\ref{fig:test_case2_hv}. 
This might happen due to bad selection of initial guesses for BiMADS. Moreover, LHS, being a statistical method, might occasionally identify Pareto optimal designs faster than deterministic algorithms. It can be observed from Figure~\ref{fig:test_case2_hv} that, after achieving a certain hypervolume space in the beginning of exploration process, LHS does not improve the Pareto frontier approximation significantly further. In contrast, BiMADS, having a poor initial guess, improves the  solution and outperforms LHS when reaching an evaluation limit.

Consider Table~\ref{tab:testcase2}, which lists designs identified by BiMADS. Similarly to the software implementation of the fast gradient algorithm, the weight matrix parameter $q_{ratio}$ is the same for all designs and equal to the lower bound for this parameter. Other design parameters, including the hardware parameter $N_{frac}$, are changing along the Pareto frontier, trading off resource usage against performance.
\begin{table}
\caption{Pareto-frontier approximation identified by BiMADS for FPGA implementations of FGM (corresponds to Figure~\ref{fig:test_case2}). Designs are sorted according to their settling times in ascending order.}
\centering
  \begin{tabular}{ c*{7}{c}}
    \hline \hline \noalign{\vskip 0.03in}
    \multicolumn{5}{c}{Design parameters} &	\multicolumn{2}{c}{Design objectives}\\
    $T_s$ & $N$ & $N_{FGM}$ & $q_{ratio}$ & $N_{frac}$ & \makecell{Settling \\ time [s] } & \makecell{Alg. \\ time [s]} \\ \hline  
   	0.2132  &  9 &  230 &  0.20 &  25 & 8.3175  & 0.1321   \\
    0.2347  &  6 &  62  &  0.20 &  25 & 8.3329  & 0.1077   \\
    0.1870  &  4 &  40  &  0.20 &  17 & 8.7067  & 0.0596   \\
    0.3802  &  4 &  83  &  0.20 &  17 & 9.0659  & 0.0559   \\
    0.3802  &  4 &  62  &  0.20 &  17 & 9.1092  & 0.0558   \\
    0.5000  &  1 &  55  &  0.20 &  17 & 10.5620 & 0.0488   \\
    0.5000  &  1 &  34  &  0.20 &  16 & 10.6264 & 0.0467   \\
    0.4141  &  1 &  20  &  0.20 &  15 & 10.6878 & 0.0462   \\
    0.4141  &  1 &  20  &  0.20 &  13 & 12.4817 & 0.0396   \\
    0.5000  &  1 &  20  &  0.20 &  13 & 13.5297 & 0.0396   \\
    0.4785  &  1 &  20  &  0.20 &  13 & 13.8890 & 0.0396   \\ \hline \hline \noalign{\vskip 0.03in}
  \end{tabular}
\label{tab:testcase2}
\end{table}

Given the Pareto frontier (Figure~\ref{fig:test_case2}) a designer will be able to select a particular implementation based on the available FPGA resources ($R_{FPGA}$). For example:
\begin{itemize}
	\item For $R_{FPGA} = 0.1$, LHS-based exploration achieves a settling time of 10.87 s, while optimization-based design allows settling of the plant in 8.71 s.
	\item Tightening the resource limit to $R_{FPGA} = 0.05$ leads to sampling times of 19.31 s and 10.56 s for LHS- and optimization-based approaches accordingly.
\end{itemize}

\section{Summary and future work} \label{chapter5_summary}

This paper proposed automating predictive control design by employing systematic optimization. Several contradicting design objectives were discussed and a bi-objective optimization problem formulation was proposed. Decision variables for the resulting optimization problem include both software and hardware parameters, which allows simultaneous optimization of both software and hardware and hence exploiting coupling between these components. It was shown that the bi-objective optimization-based design outperforms conventional exploration techniques and allows systematic investigation of resource-performance trade offs.

Future work includes exploiting the coupling between the nature of the design problem and design optimization algorithm. For instance, closed-loop cost function properties, which were recently studied in~\cite{mpc_value_article}, can be used for building and updating a surrogate model and hence reducing the number of required simulations~\cite{nomad}. Another direction for further research is deriving analytical bounds for objective functions in order to eliminate unpromising designs without actual function evaluations.


\balance

\begin{appendices}

\section{Experimental setup data} \label{sec:appendix1}

This appendix presents missing data for the experimental setup description.

Mass-spring-damper system parameters, starting from the mass attached to the wall:
\begin{center}
\begin{tabular}{ c*{2}{c} } \hline \hline \noalign{\vskip 0.03in}
\shortstack{Mass, \\ kg} & \shortstack{Spring constant, \\ N/m} &   \shortstack{Damping constant, \\ N $\cdot$ s/m} \\ \hline
0.1966  & 0.5472 & 0.0509 \\
0.2511  & 0.1386 & 0.1629 \\
0.6160  & 0.1493 & 0.0487 \\
0.4733  & 0.2575 & 0.1859 \\
0.10517 & 0.08407& 0.0700 \\
0.4509  & 0.1966 & 0.5472 \\
0.1629  & 0.0511 & 0.0386 \\
0.1487  & 0.6160 & 0.1493 \\
0.1859  & 0.4733 & 0.2575 \\
0.0700  & 0.10517& 0.08407\\ \hline \hline
\end{tabular}
\end{center}

Initial conditions for closed-loop simulations, obtained with a random number generator with the range  $ [  -0.2, 0.2  ]$:
\begin{center}
\begin{tabular}{ c*{4}{c} } \hline \hline  \noalign{\vskip 0.03in}
\shortstack{Initial \\ state 1} &  \shortstack{Initial \\ state 2} & \shortstack{Initial \\ state 3} & \shortstack{Initial \\ state 4} & \shortstack{Initial \\ state 5} \\ \hline
 0.0329 &    0.0163 &    0.1480 &   -0.0941 &   -0.0728 \\
-0.1523 &    0.1759 &    0.0582 &   -0.0082 &    0.0557 \\
0.0179  &    0.0589 &    0.0176 &    0.0884 &    0.0090 \\
0.1975  &   -0.1125 &   -0.1577 &   -0.1561 &   -0.1746 \\
-0.0382 &   -0.0207 &   -0.0537 &    0.1054 &    0.0512 \\
0.1088  &    0.1731 &    0.1891 &   -0.1232 &   -0.1445 \\
0.0785  &   -0.1625 &    0.0102 &    0.0121 &    0.1445 \\
-0.0061 &   -0.0426 &    0.0686 &    0.0965 &    0.0080 \\
-0.0609 &   -0.1400 &    0.0344 &   -0.0951 &   -0.1822 \\
0.1020  &   -0.1029 &   -0.0230 &    0.0751 &   -0.0563 \\
0.0945  &   -0.0421 &    0.0734 &    0.0816 &   -0.0231 \\
-0.1922 &   -0.0677 &   -0.0303 &   -0.0919 &   -0.1212 \\ 
0.1287  &   -0.0280 &    0.1551 &   -0.0435 &    0.1076 \\
-0.0413 &    0.1234 &    0.1020 &   -0.0490 &   -0.1136 \\
0.1162  &    0.1797 &   -0.0690 &    0.0685 &   -0.0245 \\
0.1334  &    0.1075 &   -0.1331 &    0.1448 &    0.1959 \\
0.0058  &    0.1537 &    0.0352 &   -0.1381 &   -0.1201 \\
-0.0372 &    0.0995 &    0.1302 &    0.1160 &   -0.0726 \\
0.0136  &   -0.1640 &   -0.1553 &   -0.1455 &    0.0715 \\
-0.0019 &   -0.1241 &   -0.0020 &   -0.1410 &   -0.1780 \\ \hline \hline
\end{tabular}
\end{center}

\end{appendices}

\bibliographystyle{unsrt} 
\bibliography{references}

\end{document}